\documentclass[12pt]{iopart}
\usepackage{graphicx}
\usepackage[defs]{ams}
\usepackage{url}
\usepackage{epstopdf}

\usepackage{color}
\newcommand{\red}[1]{  \textcolor{black}{#1}}
\newcommand{\blue}[1]{  \textcolor{black}{#1}}

\begin{document}

\title{Evidence for semiconducting behavior with a narrow band gap of Bernal graphite}

\author{N. Garc\'ia}
\address{Laboratorio de F\'isica de Sistemas Peque\~nos y
Nanotecnolog\'ia,
 Consejo Superior de Investigaciones Cient\'ificas, E-28006 Madrid, Spain.}
 \address{Laborat\'orio de Filmes F\'inos e Superf\'icies, UFSC, Florian\'opolis,
 Brazil.}
\author{P. Esquinazi$^\star$, J. Barzola-Quiquia, and S. Dusari }
\address{Division of Superconductivity and Magnetism, Institut
f\"ur Experimentelle Physik II, Universit\"{a}t Leipzig,
Linn\'{e}stra{\ss}e 5, D-04103 Leipzig, Germany}

% Place your abstract within the special {sciabstract} environment.

\begin{abstract}

We have studied the resistance of a large number of highly
oriented graphite samples with areas ranging from several mm$^2$
to a few $\mu$m$^2$ and thickness from $\sim 10~$nm to several
tens of micrometers.  The measured resistance can be explained by
the parallel contribution of semiconducting graphene layers with
low carrier density $< 10^9$~cm$^{-2}$ and the one from
metalliclike internal interfaces. The results indicate  that ideal
graphite with Bernal stacking structure is a semiconductor with a
narrow band gap $E_g \sim 40~$meV.

\end{abstract}

\pacs{73.90.+f,61.80.-x,81.05.Uw}
\vspace{3cm}
$\star$e-mail: esquin@physik.uni-leipzig.de\\
\submitto{\NJP} \maketitle

\section{Introduction}

The study of the transport properties of graphite has been object
of discussion for more that 60
years\cite{wal47,haewal,hae60,sw58,mcc1,mcc2}. The work of
Slonczewski and Weiss  \cite{sw58} as well as that of McClure
\cite{mcc1,mcc2} - both  a continuation of the work of Wallace
\cite{wal47} on the band structure of graphene - introduced
several (free) coupling parameters within a tight binding
calculation and the k.p. method to obtain the graphite band
structure. The parameters were then fixed using as reference
transport results obtained in bulk graphite samples with  carrier
density  per graphene layer
 $n(T\rightarrow 0) = n_0 \gtrsim 10^{10}~$cm$^{-2}$.
  The apparent existence
of a relatively large carrier density seemed to be compatible with
the relatively small and metalliclike resistivity measured in
several bulk graphite samples \cite{kelly}. All these early
results obtained for the graphite structure were resumed later
within the so-called two-bands model for graphite developed by
Kelly \cite{kelly}. After that, many band structures calculations
using the local density as well as others approximations \cite{sch92}
 reached the same conclusions and claimed to fit well the
experimental results,  see e.g. Refs.~\cite{gru08,sch09}. Note,
however, that all these calculations depend on parameters that are
being fixed to fit experimental data. If those experimental
results do not reflect the {\em intrinsic} properties of ideal
graphite, clearly those calculations as well as the obtained
parameters cannot be taken as intrinsic of the graphite structure.

A different approach to obtain the binding energy between the
planes of graphite has been proposed using Lennard-Jones potential
and assuming that the binding between graphene layers is due to
van der Waals forces \cite{gir00,gir56,gar11-vdw}. These forces
represent a very weak interaction and therefore it is rather clear
that this approach will not give  a similar band structure for
graphite as the one proposed with large coupling between the
graphene planes. From the experimental point of view we may now
doubt  that the large amount of the reported data, e.g. $n(T)$, do
reflect ideal graphite. The exhaustive experience accumulated in
gapless or narrow gap semiconductors \cite{tsi97} indicates
already that care should be taken with the measured carrier
density because it can easily be influenced by impurities and/or
defects in the graphite/graphene lattice\cite{arn09,sta07}.

We believe that the multigrain morphology of real oriented
graphite samples has not been taken into account appropriately in
the literature, especially its influence to the carrier density
and other transport phenomena. For example, electron back
scattering diffraction (EBSD) reveals  typical size of the grains
of a few microns in the $a,b$ plane within  HOPG samples, see
Ref.~\cite{gon07}, a size that limits the carrier mean free path
at low temperatures \cite{gar08}. Moreover, the contribution to
the transport properties of internal two-dimensional interfaces
found recently between single crystalline regions inside the HOPG
samples \cite{bar08} was apparently never taken into account in
the existence literature of graphite. Why these interfaces can be
of extreme importance in graphite?  As the results from various
semiconductors show (e.g. n-Ge bicrystals \cite{vul79,uch83},
p-InSb \cite{her84} as well as in Hg$_{1-x}$Cd$_x$Te grain
boundaries \cite{lud92}) internal interfaces lead to the formation
of confined quasi-two dimensional carrier systems with $n_0 \sim
10^{12} \ldots 10^{13}~$cm$^{-2}$ and clear signs for the Quantum
Hall effect\cite{uch83}. Recently done studies demonstrated the
large sensitivity of the resistivity of graphite samples to the
internal interfaces that exist between crystalline regions of
$\sim 30 \ldots 100~$nm thickness, a few microns long and running
mostly parallel to the graphene planes\cite{bar08}. These results
indicate that the earlier reported values of $n(T)$ as well as the
metalliclike behavior of the resistivity $\rho(T)$ are {\em not}
intrinsic of the graphite Bernal structure but are due to a large
extent to the contribution of internal interfaces and defects.

The aim of this work is to propose a simple model to explain the
experimental longitudinal resistance data obtained in different
oriented graphite samples of different thickness and areas.  The
rather complicated behavior of the longitudinal resistivity can be
explained assuming the parallel contribution of regions with
semiconducting graphene layers and the ones from the interfaces
between them, as transmission electron microscopy studies
revealed\cite{bar08}.

\section{Sample characteristics}

\begin{figure}[]
%\vspace{0.5cm}
\begin{center}
\includegraphics[width=1\columnwidth]{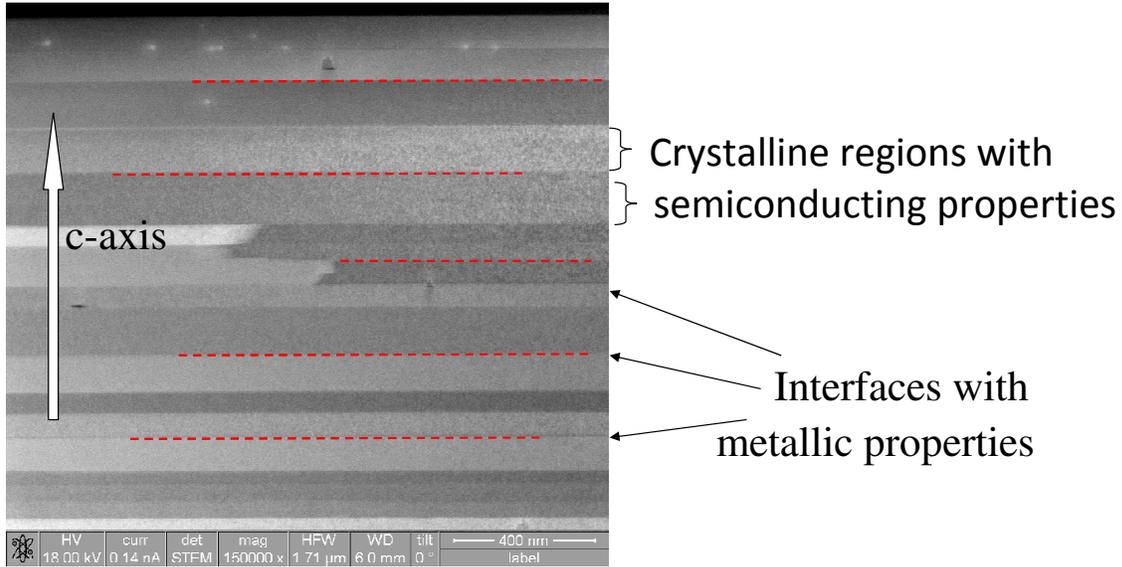}
\caption[]{Transmission electron microscope picture of the internal microstructure of
a highly oriented pyrolytic  graphite sample as shown in Ref.\cite{bar08} and similar to those
 used in this study. The dashed red lines indicate some of the
interface regions between the single crystalline graphite parts. These interfaces run parallel to the
graphene layers up to several micrometers till a larger defect is encountered (see the
defective region in the middle of the picture). Regions with different gray colors indicate slightly different
orientation of the graphite structure. From measurements with EBSD we know that the usual size of these
single crystalline regions in the $a,b$ plane is less than 10$~\mu$m \cite{gon07,gar08}.} \label{pic}
\end{center}
\end{figure}

All the graphite samples were taken from the same Bernal-type
highly oriented pyrolytic graphite (HOPG) sample from Advanced
Ceramic \red{of high purity (see,
e.g., Ref.~\cite{ohldagnjp})} and with a rocking curve width of $(0.35 \pm 0.1)^\circ$.
\blue{As shown by Bernal\cite{ber24}, this structure has the usual ABABABA$\ldots$ stacking of the single graphene layers
inside graphite. This is the stable graphite structure, which is obtained by suitable annealing treatments eliminating
also possible rhombohedral modification that results from deformation of the original hexagonal structure.}
The internal structure of the
used samples is shown in a transmission electron microscope picture in
Fig.~\ref{pic}. As shown in Ref.~\cite{bar08} this picture reveals single crystalline
regions of graphene layers of thickness between 30~nm and $\sim 100~$nm. \red{These
regions are clearly recognized in Fig.~\ref{pic} through the different gray scales and have
sightly different orientation between each other, e.g. different angle misalignments.
We note that a rotation up to 30 degrees between the graphene layers from
neighboring graphite regions has been seen by HRTEM in few layers graphene sheets in
Ref.~\cite{war09}.}

Important to understand the measured
behavior of the resistance of graphite samples is the existence of well defined interfaces
between the single crystalline regions. As we mentioned in the introduction, interfaces between
 crystalline regions in  semiconductors
with different orientations for example \cite{vul79,uch83,her84,lud92}, lead to confined quasi-two dimensional
carrier systems with much larger carrier density than the bulk matrix. We assume therefore that
 these interfaces running parallel to the graphene layers of the graphite structure
are the origin of the metallic like resistivity as well as for the apparent large carrier density
measured in bulk samples, i.e. $n_0 \gtrsim 10^{10}~$cm$^{-2}$, an assumption that is supported
by the change of the absolute resistivity with thickness \cite{bar08}.

Taking into account the internal microstructure of the graphite
samples it is clear that it is necessary to measure the resistance
of samples of small enough  thickness in order to get the
intrinsic transport of the graphite structure with its weak
coupled  graphene layers. The micrometer small samples and with
thickness below 100~nm were prepared by a rubbing
method\cite{bar08}. The resistivity data were obtained with four
contacts and checking the ohmic behavior at room temperature. The
deposited Pd/Au contacts on the micrometer large graphite flakes
were prepared by electron lithography and micro-Raman
characterization was used to check for the sample
quality\cite{bar08}, as the inset in Fig.~\ref{nT-MG}
demonstrates. Following the characterization work of
Ref.~\cite{fer06}, the Raman spectra confirm the similar stacking
between the bulk and thin graphite samples. Micrometer large mean
free path at 300~K and low carrier density\cite{dus11,esq11} are a
further proof of the high quality of the thin graphite flakes
discussed in this work.

\section{Temperature dependence of the carrier density}

We start with the  temperature dependence of the carrier density
$n(T)$ obtained from a 40~nm thin and several micrometers large
graphite flake sample. It is important to recognize that due to
the large mean free path of the carriers $\ell$ in the graphene
layers within the graphite structure, ballistic not diffusive
transport has to be taken into account\cite{gar08,esq11}. To obtain the
mean free path or the intrinsic carrier density of the graphene
layers inside graphite one cannot use straightforwardly the
Boltzmann-Drude approach to interpret the longitudinal and
transverse resistances. Due to the large mean free path of the
carriers in the weakly coupled graphene layers within the graphite
structure it is possible to use experimental methods where the
ballistic transport is clearly revealed. One possible experimental
method is the constriction method based on the measurement of the
longitudinal resistance $R$
 as a function of the width $W$ of a constriction located between the voltage electrodes.
 When $\ell \gtrsim W$ the ballistic contribution overwhelms the diffusive ones and
 allows  us to obtain $\ell(T)$ and $n(T)$ without the need of free
parameters or arbitrary assumptions\cite{gar08,dus11}. Other experimental method is the
measurement of the length dependence of the resistance. If ballistic transport is
important then a finite resistance is extrapolated at zero sample length, a value that
can be used to obtain directly the mean free path without free parameters or arbitrary
assumptions \cite{esq11}. Both independent methods provide similar large
carrier mean free path as well as much lower carrier density as those
found in the literature for bulk graphite samples.

\begin{figure}[]
%\vspace{0.5cm}
\begin{center}
\includegraphics[width=0.8\columnwidth]{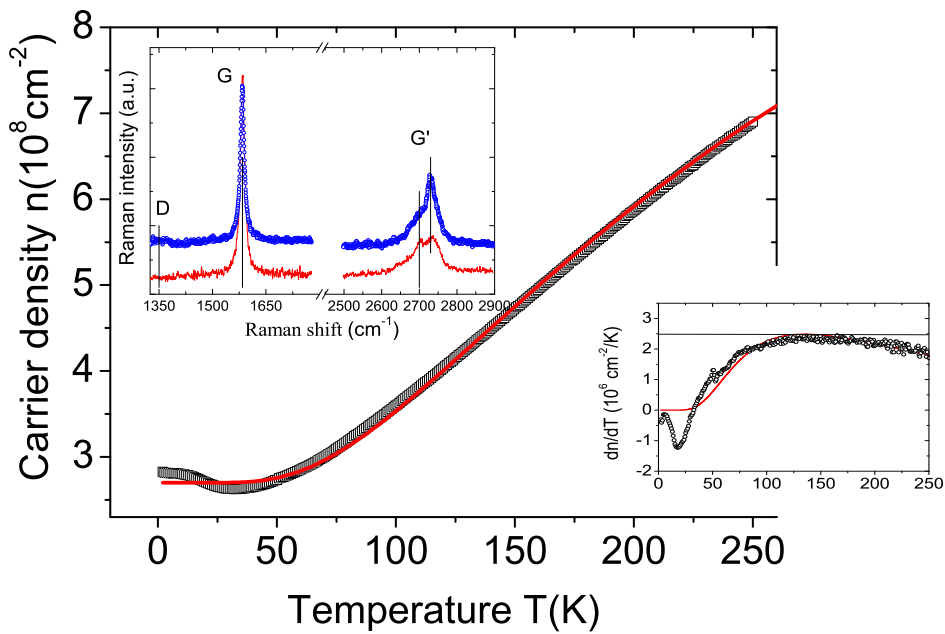}
\caption[]{Carrier density per graphene layer obtained using the
constriction method \protect\cite{gar08} for a graphite sample of
size $ 9 \times 3 \times 0.040~\mu$m$^3$. The continuous line is a
fit to the data and follows $n(T)[10^8$cm$^{-2}] = 2.7 + 12.4
\exp(-540/2T[$K$])$. We estimate a $\sim 30\%$ error  in the
absolute value of the carrier density, mainly due sample geometry
errors as well as in the constriction widths. The upper left inset shows the
Raman (514~nm) spectra of: $(\circ)$ bulk graphite (Fig.2(a)) and
a (red line) multigraphene sample (similar to that of Fig.3(a)).
The absence of a D-peak indicates the absence of a significant
number of defects. \red{The bottom right inset shows the temperature dependent
derivative of the carrier density (circles) and
of the fitting curve shown in the main panel. Note the maximum in the derivative at $\sim 125~$K present
in both derivatives. The horizontal straight line is only a guide to the eye.}} \label{nT-MG}
\end{center}
\end{figure}

Figure~\ref{nT-MG} shows $n(T)$ for a thin graphite sample; it
 follows an exponential dependence with an
activation energy or energy gap $E_g \sim 46~$meV.
\red{Due to the small gap and from Fig.~\ref{nT-MG} it is not obvious
that the experimental curve follows an exponential behavior and not a linear one at
temperatures above the saturation region. A simple
way to distinguish that, it is calculating the temperature derivative. In case of a linear temperature behavior
we expect a saturating upper derivative value at high temperatures  but a shallow maximum for the exponential function
at intermediate temperatures. The bottom right inset shows the calculated derivative for
both curves, the experimental and the exponential function fit. It is clear that the exponential function
provides the correct temperature dependence of the carrier density.}
The shallow
minimum in $R(T)$ (main panel in Fig.~\ref{nT-MG}) at $T < 50~$K is an artifact due to interfaces, at least
that one between the substrate and sample and/or the sample free
surface.  The finite value of $n(T \rightarrow 0)$ can be due to
this interface contribution or due to lattice defects\cite{arn09}
including a very small amount of impurities like hydrogen; note
that $n = 10^8~$cm$^{-2}$ would mean of the order of a single
hydrogen atom or C-vacancy in $1~\mu$m$^2$ graphene area.

One may be surprised to get such low carrier density in comparison
with usual values reported in the literature for graphite, e.g.
$n_0 \gtrsim 10^{10}~$cm$^{-2}$ \cite{kelly,gru08,sch09}. We note
first that the measured graphite sample has a thickness less than
50~nm and therefore much less contribution of interfaces. Second,
the measured carrier density of graphite reported in literature
was obtained mostly for bulk samples with an unknown concentration
of interfaces as well as defects. Further independent evidence
that supports an intrinsic low carrier density in graphite is
given by the vanishing of the Shubnikov-de Haas (SdH) oscillations
in the magnetoresistance the thinner the graphite sample, an
experimental fact already reported in 2001 that to our knowledge
remained without explanation \cite{oha01}. Note that older
publications reported SdH oscillations of very large amplitude,
e.g. Ref.~\cite{sou64}, in comparison with the rather weak
amplitudes, if at all observed, found nowadays in better quality
samples or thin enough graphite samples. A clear demonstration
that these SdH oscillations are actually non-intrinsic of ideal
graphite is given by their appearance in thin graphite samples
after introducing defects by irradiation as shown in
Refs.~\cite{arn09,barnano10}.

\section{Temperature dependence of the resistance of different graphite samples}

Following the results from Ref.~\cite{bar08}  as well as the semiconducting behavior of
$n(T)$, see Fig.\ref{nT-MG},  obtained for a thin graphite sample, we assume therefore that
the graphene layers inside each graphite sample are semiconducting
and altogether and between the voltage electrodes their signal depends on
an effective resistance of the type:
\begin{equation}
R_s(T) = a(T) \exp(+E_g/2k_BT)\,. \label{rexp}
\end{equation}
The prefactor $a(T)$ depends basically \red{on the mobility, i.e. the mean free path and
 on details of the carriers band
structure (e.g. effective mass).  If we take into account in $a(T)$  the temperature dependence of
the mean free path  recently obtained for similar
samples, i.e. $\ell(T) \simeq ((3)^{-1} + (6.4 \times 10^5/T^2)^{-1})^{-1}$ ($\ell$ in $\mu$m and $T$ in K), see
Ref.~ \cite{esq11},  the absolute value of
the energy gap $E_g$ obtained
from the fits to the data  remains the same within the
confidence limits of 35\%. For simplicity
we will take  $a(T)$ as a temperature independent parameter as
well as the energy gap $E_g$.} The absolute value of the prefactor $a$ (as well as of the
other prefactors)
will change from sample to sample; note that we estimate  the
resistance of a given sample not its resistivity.

For samples with thickness larger than $\sim 50~$nm and
of several micrometers length there is a larger probability to have interfaces, which signals will be
picked up by the voltage contact electrodes (usually several micrometers apart).
Therefore, in parallel to $R_s(T)$ we simulate the contribution from the
interfaces through the resistance:
\begin{equation}
R_i(T) = R_0 + R_1 T + R_2 \exp(-E_a/k_BT)\,, \label{rsup}
\end{equation}
where the coefficients $R_1, R_2$ as well as the activation energy
$E_a$ are free parameters. The temperature independent term in
Eq.~(\ref{rsup}) represents the residual resistance measurable at
low enough temperatures. The  unusual thermally activated term as
well as the, in general weak, linear one will be discussed at the
end. The total resistance $R_T(T)$ is given by the parallel
contribution of $R_s$ and $R_i$ as  $R_T(T) = [R_s(T)^{-1} +
R_i(T)^{-1} ]^{-1}$. Clearly, by changing the parameters one can
obtain all types of behavior for $R(T)$. We will see that  a
consistent description of the data can be indeed reached and that the
main free parameter $E_g$ is similar for all samples.

\begin{figure}[]
%\vspace{0.5cm}
\begin{center}
\includegraphics[width=0.33\columnwidth]{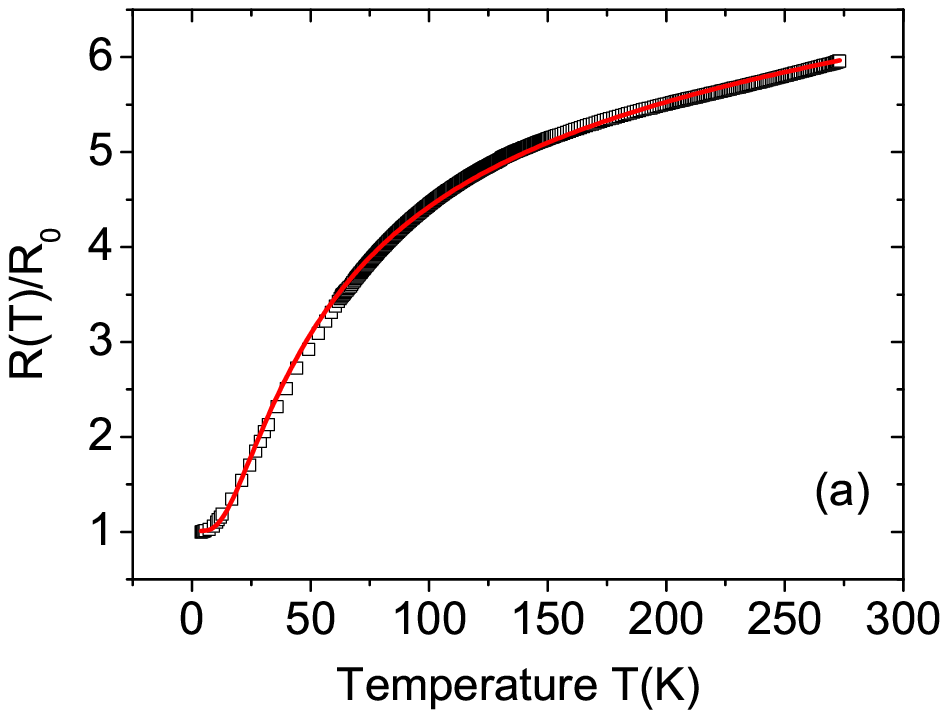},\includegraphics[width=0.33\columnwidth]{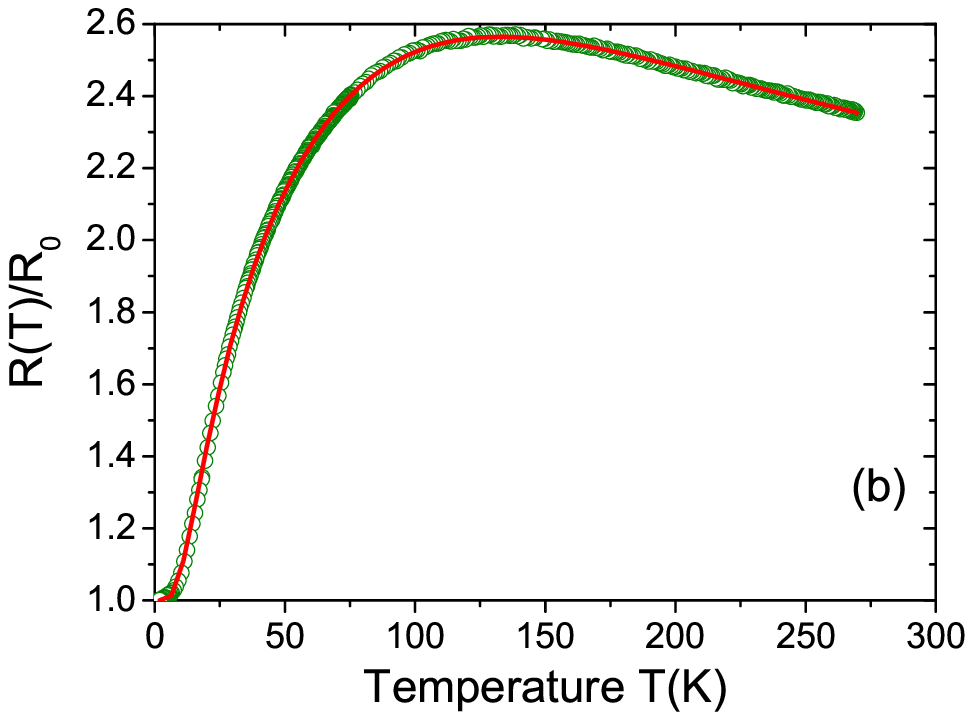},\includegraphics[width=0.33\columnwidth]{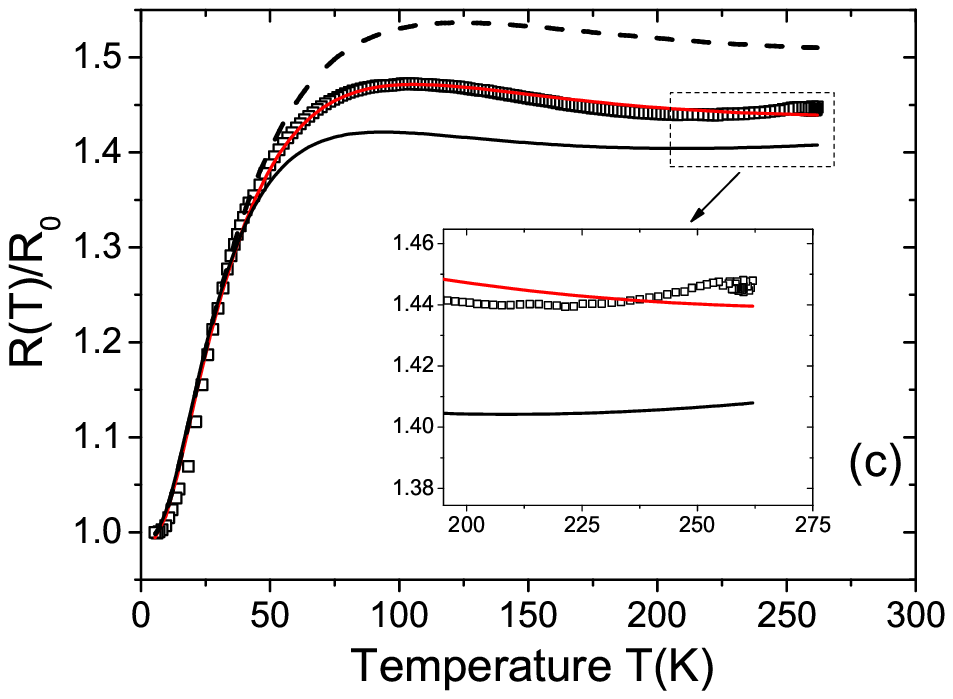}
\caption[]{Normalized resistance $R/R_0$ vs. temperature for three
 HOPG samples with thickness, length (between voltage electrodes), width and $R_0$:
 (a) $\simeq 20~\mu$m, 2~mm, 1~mm,  $0.003~\Omega$;
 (b) $\simeq 10~\mu$m, 1~mm, 1~mm, $0.013~\Omega$;
(c) $\simeq 50~$nm,  $\sim 3~\mu$m,  $\sim 3~\mu$m, $15~\Omega$.
 The (red) lines through
the experimental data are obtained from the the parallel
contributions given by Eqs.~(\protect\ref{rexp}) and
(\protect\ref{rsup}): (a) $R_i/R_0 = 1 + 2.2 \times 10^{-3} T[$K$]
+ 5.2 \exp(-48/T[$K$]$); (b)
 $R_i/R_0 = 1 + 2.2 \exp(-33/T[$K$])$ and $R_s(T)/R_0 = 3.4 \exp(662/2T$[K$])$;
 (c) $R_i(T)/R_0 = 1 + 2 \times 10^{-3} T[$K$] +  0.7 \exp(-38/T[$K$])$
 and $R_s(T)/R_0 = 2.35 \exp(340$/2T[K$])$. Note that the curves
  obtained are very sensitive to small
changes of parameters. For example in (c), the continuous curve
just below follows $R_i(T)/R_0 = 1 +  2.41 \times 10^{-3} T[$K$] +
0.7 \exp(-38/T[$K$])$ and $R_s(T)/R_0 = 2.35 \exp(260/2T$[K$])$.
Whereas the dashed curve above follows $R_i(T)/R_0 = 1 +  2.17
\times 10^{-3}T[$K$]  + 0.7 \exp(-38/T[$K$])$ and $R_s(T)/R_0 =
2.35 \exp(400/2T$[K$])$. The inset in (c) blows out the high
temperature part where one recognizes the slight increase of the
resistance at $T >  200~$K, an increase that can be obtained
changing slightly the used parameters as the lower curve shows.}
\label{HOPG}
\end{center}
\end{figure}

\begin{figure}[]
\vspace{-0.5cm}
\begin{center}
\includegraphics[width=0.3\columnwidth]{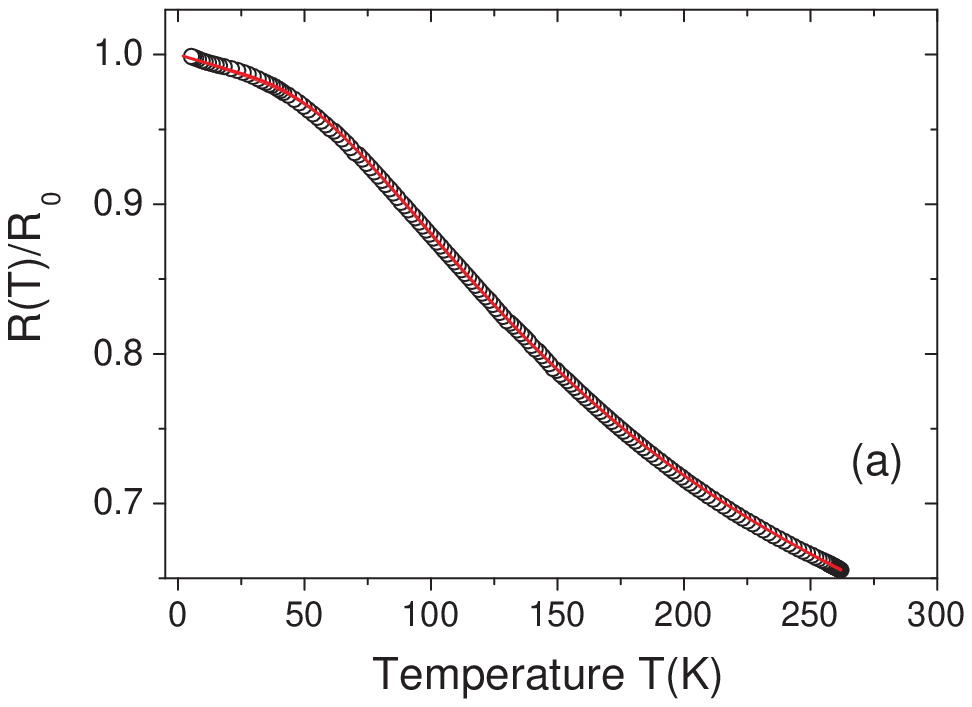},
\includegraphics[width=0.3\columnwidth]{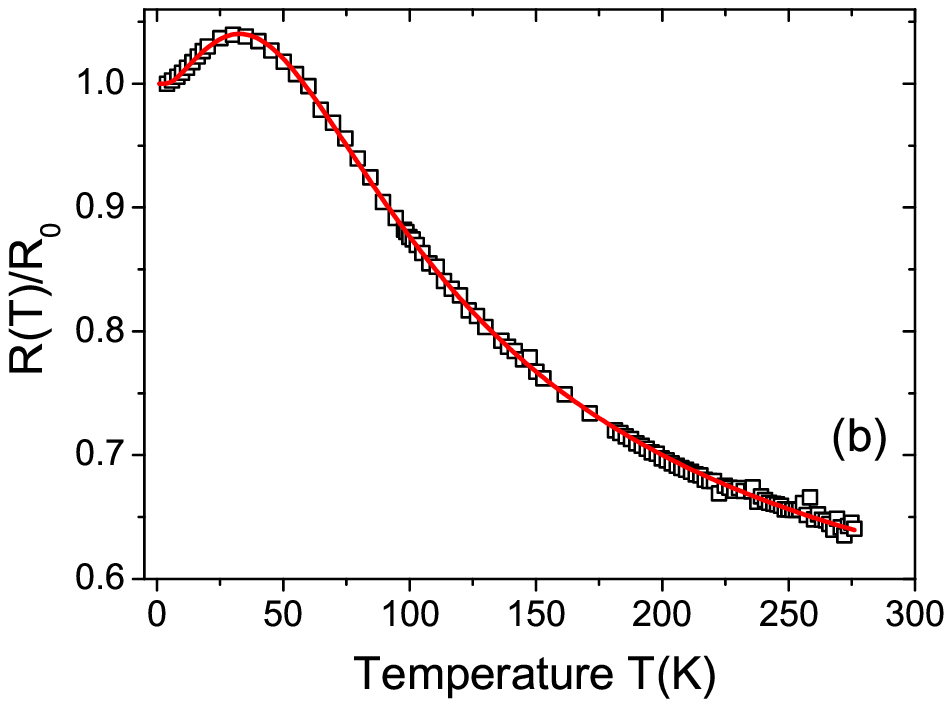}, \includegraphics[width=0.3\columnwidth]{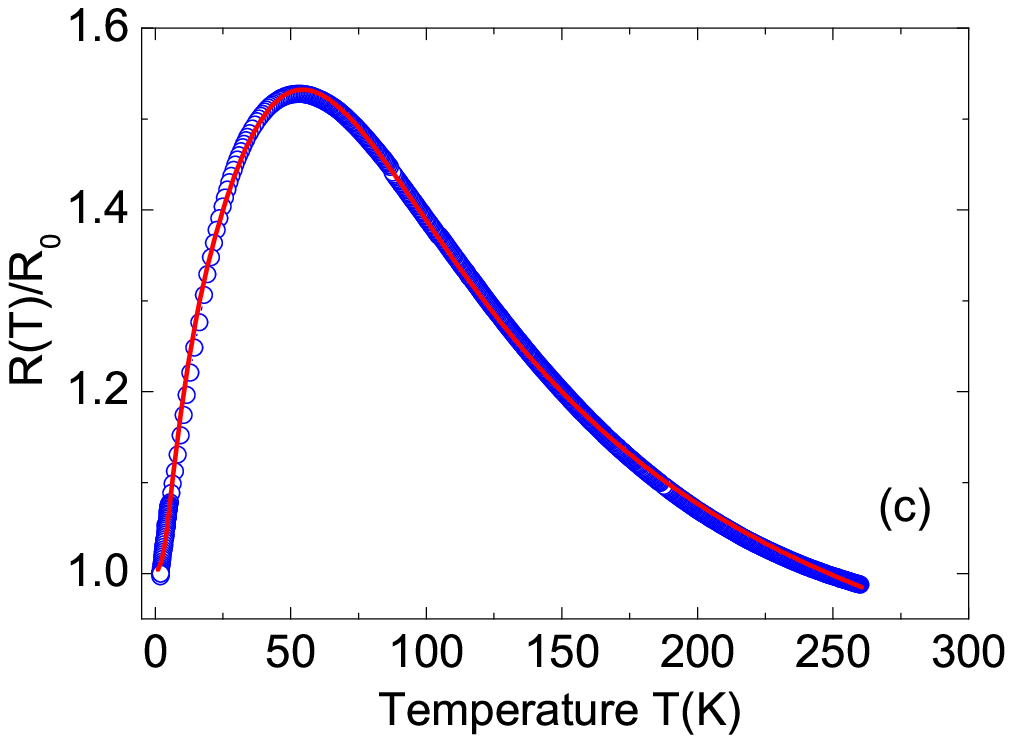}
\caption[]{Similar to Fig.~\protect\ref{HOPG} but for samples with
geometry and $R_0$: (a) $\simeq 13~$nm, $14~\mu$m~, $10~\mu$m,
$490~\Omega$; (b) $\simeq 20~$nm, $ 5~\mu$m~ , $10~\mu$m,
$32~\Omega$; (c) $\simeq 37~$nm,  $ 27~\mu$m, $6~\mu$m,
$69~\Omega$. The lines through the experimental data follow: (a)
 $R_i(T)/R_0 = 1 - 6.2 \times 10^{-4} T[$K$]$ and $R_s(T)/R_0 = 1.18 \exp(480/2T$[K$])$; (b) $R_i(T)/R_0 = 1
+ 0.096 \exp(-24/T[$K$])$ and $R_s(T)/R_0 = 0.82
\exp(350/2T[$K$])$. (c)   $R_i(T)/R_0 = 1 + 0.0044 T[$K$] + 0.47
\exp(-12/T[$K$])$ and $R_s(T)/R_0 = 0.73 \exp(400/2T[$K$]$. }
\label{MG}
\end{center}
\end{figure}

Figures~\ref{HOPG} and \ref{MG} show the normalized resistance vs.
temperature and the fits with the parallel resistance model
described above of six different samples with different weights
between the {\em intrinsic}  semiconducting and the {\em
non-intrinsic} interface contribution. That is why the pre-factors
of the main terms in $R_T(T)$  must change from sample to sample,
i.e. $R_1/R_0, R_2/R_0$ and $a/R_0$, because we do not fit the
resistivity but the resistance. \red{From all the obtained data and
within the confidence limits of the fitting we obtain an energy
gap $E_g = 40 \pm 15~$meV, independently of the sample geometry, i.e.
a similar energy gap is obtained for very thin as well as thick graphite samples if
the contribution of the interfaces does not short circuit completely that from
the graphite crystalline regions. This fact indicates that we are not dealing here
with a special graphite structure found only in a few tens of nanometers thick samples.}
This is the main result of this study.

We stress that the interfaces as well as the single crystalline
regions are restricted in thickness as well as in the $a,b$ plane
parallel to the graphene layers. This fact supports the use of the
simple parallel resistance model. For thick enough graphite
samples (thickness $t \gtrsim 1~\mu$m) the contribution of the
interfaces overwhelms and the measured resistance will be mainly
given by Eq.~(\ref{rsup}), as the results for bulk HOPG sample
shows, see Fig.~\ref{HOPG}(a). The exponential term with an
activation energy $E_a/k_B = (48 \pm 2)$K is of the same order as
reported in Ref.~\cite{yaknar} for similar samples. To fit the
data shown in Fig.~\ref{HOPG}(a)
 a relatively small linear-temperature term contributes at high temperatures.

%\begin{widetext}


%\end{widetex

In general the  semiconducting behavior starts to be clearly
visible in thinner samples due to the decrease in the amount of
interfaces; obviously  there is no sharp thickness threshold
because the properties of the interfaces as well as the defects
and carrier concentrations vary from sample to sample,
affecting the relative weight between the two parallel
contributions. The results shown in Figs.~\ref{HOPG}(b) and (c)
are examples where the two contributions $R_i$ and $R_s$ compete.
For samples thinner than $\sim 40~$nm the semiconducting behavior
starts to overwhelm that from the interfaces, see Fig.~\ref{MG}.

Note that this semiconducting behavior is observed for a range of
about five orders of magnitude in sample area and that the width
or length of the samples overwhelms the intrinsic Fermi wavelength
of the graphene planes inside the sample\cite{dus11} $\lambda_F
\lesssim 1~\mu$m. Therefore, the semiconducting behavior is not
due to a quantum confinement shift, as observed in Bi
nanowires\cite{hub09} of small diameter $d < \lambda_F$. We stress
 that our results should not be interpreted as  a  metallic to semiconducting
transition as a function of thickness, but only the weight between
the two parallel contributions changes upon sample thickness.
Because at least the interfaces sample-substrate and
sample-surface remain, as for the sample of Fig.~\ref{MG}(a), both
parallel contributions are always present. We note that the
$T-$dependence of the $c-$axis resistivity of graphite with
pressure \cite{uhe87} shows similar characteristics as the ones
for the resistivity in plane shown here. The $c-$axis resistivity
behavior was basically interpreted with a tunnel barrier for the
electrons between planes and crystallites.

\section{Discussion}

We would like to discuss the question of the existence of
an energy gap taking into account results from literature and whether
 this intrinsic semiconducting behavior has been observed in the
 past. (a)
As we discussed above and taking into account the contribution of
the interfaces bulk graphite samples have, it is clear
that the behavior obtained in transport measurements of those samples
is not intrinsic. Therefore we cannot consider those experimental results
as a proof for  the apparent metallicity of graphite.

(b) One would expect that
infrared (IR) measurements show somehow the semiconducting behavior. However, those
measurements, as for example in Ref.~\cite{li06}, were  done in bulk
samples (several mm$^2$ area and several hundreds of micrometers thick)
that show clear SdH oscillations \cite{yakovadv03} with the usual carrier
density above $10^{10}~$cm$^{-2}$ per graphene layer and therefore do not
represent intrinsic graphite. Note that the internal interfaces, see Fig.~\ref{pic},
and due to their large carrier density in comparison with the crystalline regions, are those
regions that provide a shielding to the electromagnetic wave used in IR measurements.
In general we remark that spectroscopy methods are not completely reliable when
one wants to resolve a gap of 40~meV in samples that are inhomogeneous for the reasons discussed above.
Therefore, we believe that the intrinsic properties of the single crystalline graphite
regions are not actually seen by this technique.
\red{Also other experimental techniques, see e.g. Ref.~\cite{kos11}, would have problems to
recognize an electronic band structure with an energy gap of the order of 40~meV if the
carrier density is much larger than the intrinsic one of ideal, defect free graphite.}

\blue{(c)  We may ask about the results on graphite using
angle-resolved photoemission spectroscopy (ARPES) published
several times in the past. There are three points we should take
into account about ARPES results. Firstly, this technique is
surface sensitive, although  we may expect to see to some extent a
vestige of the bulk band edge. Second, different samples may
provide different results. A clear correlation between sample
quality and other of its characteristics with the ARPES results is
difficult because one not not always find the necessary
information in the reported studies. Finally the scanned sample
area and the energy resolution of the usual spectroscopy systems.
We know from other techniques like electron back scattering
diffraction (EBSD) \cite{gon07}
  or electron force microscopy (EFM) \cite{lu06} that homogeneous regions of graphite appear to be
  in the region of $~10~\mu$m (parallel to the planes) or smaller.
 Regarding the energy resolution we find,
 for example, in the ARPES work of Ref.~\cite{zhou061}
 a resolution of 15 meV to 65 meV,  in Ref.~\cite{zhou062}  it is  from 15 meV to
40 meV and in Ref.~\cite{lee08} 40~meV.
 This resolution appears to be too low to clearly resolve an energy gap
of the order of 30 to 50 meV in bulk graphite. A better energy resolution between 4 to 15~meV
 has been achieved in Refs.~\cite{sug07,gru08} using kish and natural graphite samples, respectively.
 Interestingly,  in Ref.~\cite{sug07} an energy gap of 25~meV between the $\pi$ and $\pi^\star$ bands
 at the K(H) point was reported whereas in Ref.~\cite{gru08} an energy gap of $37~$meV has been
 inferred from the band fits near the H point, although the used sample
 had apparently an average carrier density of $\sim 6 \times 10^{11}$cm$^{-2}$ (at 25~K).
 Clearly, higher energy resolution, smaller scanned areas and clearer sample quality aspects
 are necessary in future experiments using this technique.}

(d) In general we note that techniques that are sensitive to the surface  do
not necessarily provide the intrinsic properties of graphite. Apart from the extra doping
graphite surfaces may have due to  defects and added atoms or molecules, we
 note that the surface itself may have a
 different electronic structure as the graphene layers inside the
 Bernal structure due to the total absence of coupling
 with a graphene layer above it. Therefore, STM
 experiments on the graphite surface do not necessarily provide the intrinsic behavior
 of graphite with certainty.  \red{However, there
are STM experiments that suggest the existence of an energy gap of
the same magnitude as the one obtained in this work. This energy
gap appeared in a few graphene layers\cite{li07} from
"unidentified states" or levels that at zero magnetic field do not
coalesce,} a result supported by Hubbard calculations with a
constant potential\cite{men10}. Clearly, if a few graphene layers
shows an energy gap, there seems to be no simple reason to rule
out that graphite should not have it. Note also that in
Ref.~\cite{li07} other electronic contributions, partially due to
Dirac fermions, have been obtained at the surface. This fact
suggests that the resistance of this interface (or between the
sample and substrate) will provide a different conduction path in
parallel to the semiconducting one from the internal graphene
layers. This gives a simple explanation for the saturation in
resistance measured at low temperatures.

\red{(e) In order to compare the results of the temperature dependence of the resistance
$R(T)$ obtained in single or a few
 layers graphene samples with our results one should compare
the one obtained at the lowest carrier density possible.
This has been done in, e.g., Ref.~\cite{ska09} where one recognizes
that at the lowest carrier density reached near the neutrality point (former
Dirac point) the resistance does not behave metallic but rather semiconducting.
Increasing the carrier density only slightly with the bias voltage, the resistance
starts to show signs of metallicity\cite{ska09} . Further semiconducting behavior  in $R(T)$ has been seen in
Refs.~\cite{mor08} and \cite{tan07}.}

(f) And at last but not
least we would like to point out the recently done optical
pump-probe spectroscopy experiments on 30~nm thick graphite flake
\cite{bre09}, a sample geometry relevant for comparison with our
study.  Even for excitation densities 10 times larger than the one
of our samples, these experiments
 suggest a renormalization of the band gap by $\sim 30~$meV,
 which is nearly constant during the first picosecond. The authors in Ref.~\cite{bre09} conclude that
  carrier equilibration in graphite is similar to semiconductors with a nonzero band gap.

The interaction between graphene
planes in graphite is van der Waals type, which gives a binding
energy of $\lesssim 400~$K. That means that each graphene layer
should be very little affected and therefore the electron electron
interaction may be also active in graphite as
 also in ideal single graphene as well. This cannot be
excluded because the data obtained in the literature are for large
$n$ and therefore any gap of the magnitude we obtain here will not
be so easily measurable. Electron interactions are large and for a
small enough carrier density the expected screening will be very
weak promoting therefore the existence of an energy gap. This is
what it is observed in Monte Carlo simulations for the unscreened
Coulomb interaction in graphene with different Dirac
flavors\cite{druprl09}.

Before concluding we would like to discuss the possible origin of the
 exponential function in Eq.~(\ref{rsup}) because this is not the usual one
 expects for metals or semimetals and cannot be understood within the usual
electron-phonon interaction mechanisms also in two dimensions. We note that
this function has been already used to describe the increasing resistance of
 bulk graphite samples with temperature and speculated to be related to some
  superconducting-like behavior
 in graphite\cite{yaknar}.  We note further that a similar
dependence has been observed in granular Al-Ge\cite{sha83}, which
shows for a particular Al concentration a
semiconductor-superconductor transition practically identical to
that in Fig.~\ref{MG}(c).  The thermally activated behavior in
Eq.~(\ref{rsup}) can be understood on the basis of the
Langer-Ambegaokar-McCumber-Halperin (LAMH) model\cite{lan67,mcc70}
that applies to narrow superconducting channels in which thermal
fluctuations can cause phase slips. The value of the activation
energy $E_a$ obtained from the fits depends on the measured sample
and it is between 10~K and 40~K for a large amount of measured
samples.

Although we do not claim that this exponential function is due to
granular superconductivity embedded within the interfaces shown in
Fig.~\ref{pic}, this speculation matches however a series of
experimental hints obtained recently. We would like to mention
that the possible existence of high-temperature superconductivity
in non-percolative regions within the internal interfaces found in
highly oriented pyrolytic graphite
 has been proposed recently \cite{bar08} .
The experimental hints are quite diverse and all of them point in the same direction:
 (1) Huge magnetic field driven metal-insulator transition in bulk graphite \cite{kempa00,heb05}. This
 transition vanishes for thin enough samples because of the lack of the internal
 interfaces.
(2) Superconducting-like hysteresis loops in the magnetization of
pure HOPG samples \cite{yakovjltp00,kopejltp07}. (3) Anomalous
hysteresis in the magnetoresistance of mesoscopic graphite
samples, similar to those found in granular superconductors
\cite{esq08,sru11}. (4) Quantum oscillations in the
magnetoresistance in multigraphene samples based on
Andreev-scattering  mechanism \cite{esq08}. (5)
Superconducting-like magnetic field driven transition in the
resistance of an internal interface of a graphite flake
\cite{barjsnm10}. (6) Finally, Josephson-like $I/V$ characteristic
curves found in thin graphite  lamellas \cite{barint}.

Since most of the theoretical predictions emphasize that
superconductivity should be possible under the premise that the
carrier density per graphene area $n > 10^{13}~$cm$^{-2}$ in order
to reach $T_c > 1~$K, it is then appealing that regions at the
internal interfaces within the graphite structure may have
superconducting patches due to its much higher carrier density. We
note that  interfaces in some semimetals materials like Bi can be
superconducting \cite{mun06,mun08}.

The weak linear term in Eq.~(\ref{rsup}) provides a typical metallic
dependence to the resistance, which  may come from the metallic
regions within the interfaces. This small term is not always
necessary to get a reasonable fit to the data and can be even
negative (or may follow a weak variable range hopping dependence)
as appears to be the case for the sample-substrate and/or
sample-surface interfaces, see Fig.~\ref{MG}(a). Even in the case of total absence of
internal interfaces, as for example for thin enough samples, it is clear that the semiconducting behavior of the
intrinsic graphite regions cannot be observed at low enough temperatures due to the
parallel contributions of the surface and/or substrate/sample.

Concluding, the temperature dependence of the electrical resistance of
oriented graphite samples can be quantitatively understood
assuming the parallel contribution of  semiconducting and
 normal metallic (and/or granular superconducting) regions. The
semiconducting contribution clearly indicates that Bernal graphite
is a semiconductor with a narrow gap of the order of 40~meV. We
speculate that the existence of narrow band gap we observe in
graphite may be applicable to other semimetals. In other words we
doubt on the existence of strictly zero band gap in nature.

This work is supported by the Deutsche Forschungsgemeinschaft
under contract DFG ES 86/16-1. S.D. is supported by the Graduate
School of Natural Sciences ``BuildMoNa" of the University of
Leipzig.

\bibliographystyle{unsrt}
%\bibliography{D:/DATA/hopg/magnetic_carbon}
%\bibliography{/Users/pablo/Documents/biblib/magnetic_carbon}

\end{document}